\documentclass[preprint]{aastex}
\usepackage{epsfig}
\usepackage{emulateapj5,apjfonts}

\begin{document}
\makeatletter
\newenvironment{inlinetable}{%
\def\@captype{table}%
\noindent\begin{minipage}{0.999\linewidth}\begin{center}\footnotesize}
{\end{center}\end{minipage}\smallskip}

\newenvironment{inlinefigure}{%
\def\@captype{figure}%
\noindent\begin{minipage}{0.999\linewidth}\begin{center}}
{\end{center}\end{minipage}\smallskip}
\makeatother

\def\wdf{white dwarf}
\def\etal{et al.}
\newcommand{\chandra}{{\it Chandra}}
\newcommand{\hst}{{\it HST}}
\newcommand{\lum}{\thinspace\hbox{$\hbox{erg}\thinspace\hbox{s}^{-1}$}}
\def\rd{Di\thinspace Stefano}

\title{Quasisoft X-Ray Sources: Unusual States of Stellar-Mass
Objects, or Intermediate
Mass Black Holes?}  

\author{R.~Di\,Stefano$^{1,2}$, F.~A.~Primini$^1$, A.~K.~H.~Kong$^{1}$.
T.~Russo$^{1,2}$}
\affil{$^1$ Harvard-Smithsonian Center for Astrophysics, 60
Garden Street, Cambridge, MA 02138}
\affil{$^2$ Department of Physics and Astronomy, Tufts
University, Medford, MA 02155}

\begin{abstract} 
Chandra observations of nearby galaxies have 
revealed a number of X-ray sources characterized by 
high luminosities ($L_X > \sim10^{36}$  erg s$^{-1}$) and $k\, T$  
in the range $100 - 350$ eV.
These ``quasisoft X-ray sources'' (QSSs)
are harder than 
luminous supersoft X-ray sources (SSSs), whose characteristic temperatures
are tens of eV, but, with  little or no emission above $2$ keV,
they are significantly softer than most canonical X-ray sources.  
They are likely to include a range of  physical systems;
some may be common systems 
in unusual states:  
neutron stars or stellar-mass black holes (BHs), 
or SNRs. Others may be
accreting BHs of intermediate mass. We have analyzed {\it Chandra} data from $19$ near-by galaxies to identify
QSSs with sufficiently high count rates to allow spectral fits.
Six of these galaxies have been studied in great detail; in these we find
$89$ SSSs and $122$ QSSs.   
In this paper we present spectra for QSSs those with more than $50$ counts. 
We also use data from
six of these galaxies to
study the broadband spectral distributions of QSSs and to compare
them with those of other X-ray sources. 
Since QSSs in the Milky Way are likely to be missed by our selection procedures, we discuss other signatures by which they can be found.  
If some of the QSSs
for which we have spectra are accreting BHs, the lower
bounds for the BH mass range
from roughly $30\, M_\odot$ to $1000\, M_\odot$.
\end{abstract} 

\keywords{black hole physics --- X-rays: binaries --- X-rays: galaxies}

\section{What are quasisoft sources?} 

\subsection{Operational Definition}

The {\it Einstein} Observatory detected a small number of
X-ray sources (XRSs) with luminosities 
near $10^{37}-10^{38}$ erg s$^{-1}$, and little or no emission above $1$ keV.
When {\it ROSAT} identified approximately a dozen 
such sources in the Magellanic 
Clouds and  Galaxy and more than $15$ in M31, the 
empirical class of luminous supersoft X-ray sources (SSSs)
was established.
SSSs of lower luminosity (but greater than roughly $5 \times 10^{35}$
erg s$^{-1}$) have also been 
discovered. The effective radii of SSSs are comparable to
white dwarf (WD) radii, although in principle, SSS-like radiation can be
emitted by objects more compact than WDs.
Indeed, roughly half of
the SSSs with optical IDs contain hot WDs--e.g., recent novae,  
symbiotics, a planetary nebula.  
The other SSSs are binaries which are thought to contain WDs
accreting at high enough rates ($\sim 10^{-7} M_\odot$ yr$^{-1}$)
to allow nuclear burning (van den Heuvel et al.\, 1992). 

Because absorption can readily hide emission from SSSs,
we can study large 
populations of them only by observing external galaxies
located along directions with small gas columns.
Unfortunately, however, their great distance from us ensures that  
we receive few photons from most XRSs in external galaxies.
It is therefore important to develop methods for identifying SSSs on the basis
of broadband fluxes rather than detailed X-ray spectra.
We have developed an algorithm
based on $3$ energy bands\footnote{S: 0.1-1.1 keV; M: 1.1-2 keV,
H: 2-7 keV}  to select SSSs
from among all of the XRSs in an external galaxy.
For details of the algorithm and
tests on simulated data, see \rd\ \& Kong (2003b); the flow chart
of Figure~1 sketches its broad outlines. 

The first $2$ steps in the algorithm
select the softest sources, those most likely to be like the  
SSSs observed in the Galaxy and Magellanic Clouds.
We refer to the sources selected by these two steps as 
``classical" supersoft
sources, or simply SSSs.  
An additional $7$ steps relax the 
selection criteria in
a hierarchical fashion, 
to avoid missing potentially important sets of very soft sources.
For example, when we observe a hot ($\sim 100$ eV) SSS located behind a
large gas column,  the relative fraction of counts recieved above $1.1$ keV might
be so large that restrictive criteria would fail to
identify the source
as an SSS, even though the unabsorbed spectrum would
fit the SSS criteria. In fact, such a hot SSS would be particularly
interesting, because it could be among the hottest of
nuclear-burning WDs and a good candidate for a possible future 
Type Ia supernova.
Less restrictive conditions also allow us to
identify as very soft, those SSSs producing so few counts that
their hardness ratios are not well constrained. 
In fact, since low-luminosity
SSSs are expected to be more numerous than those with high 
luminosities, and since absorption can decrease the
count rate significantly, many genuine SSSs we observe in other galaxies
will fall in this category.   
We therefore need to loosen the selection criteria in a systematic
way that does not rely on only hardness ratios. 
Less restrictive conditions, however, have the potential to cast
a wider net, selecting sources that are actually harder than 
the locally-studied SSSs.
In this paper,  we concentrate on the characteristics
of those sources identified by the additonal $7$ steps in our algorithm.
Because they can be hotter than the known SSSs, we
have dubbed them ``quasisoft'' sources (QSSs).  

Thus, QSSs are an operationally-defined set of sources whose distinguishing characteristics may be summarized as follows:
(1) their broadband 
 spectra should have 
no significant emission above $2$ keV, with  
$H/\Delta\, H < 0.5,$ where $H$ is the number of counts between $2$
and $7$ keV, and $\Delta\, H$ is the one$-\sigma$ uncertaintly, 
including both the effects of 
counting statistics and ambient background; or 
(2) if $H/\Delta\, H>0.5,$ there is a strict limit on the
ratio $H/T,$ where $T$ is the total number of photons,
combined with requirements on the relative significance of the detections in
the $S, M,$ and $H$ bands.
For example, one of three such requirements in our algorithm 
 is: $H/T< 0.05,$ combined with
$S/\Delta\, S > 3.$

\vspace{.1 true in}
\begin{inlinefigure}
\psfig{file=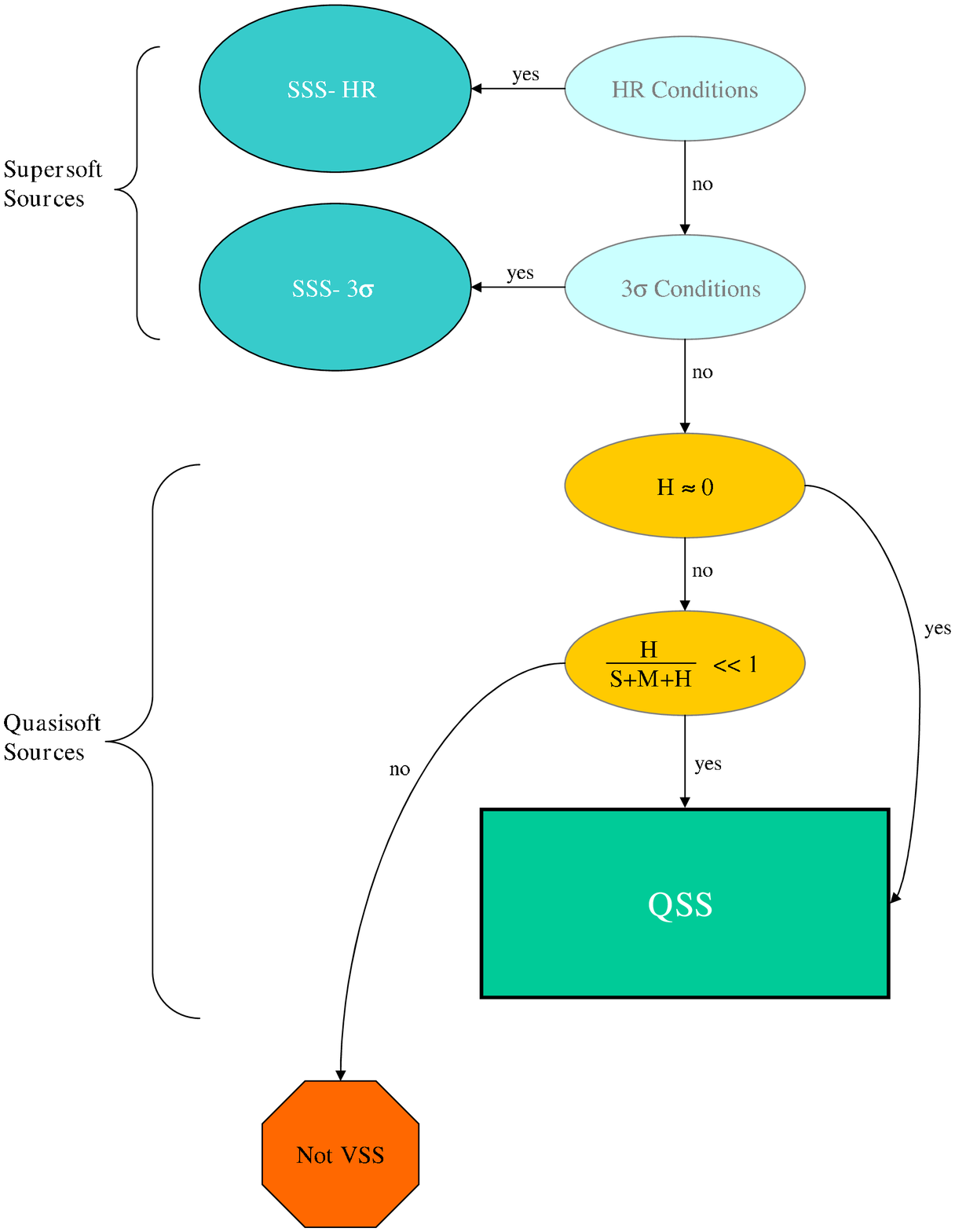,height=4.8in}
\caption{
Flow chart of the procedure by which SSSs and QSSs are selected.
Sources that satisfy the HR or $3\sigma$ conditions are referred to
as SSSs. Any source that does not qualify as an SSS passes through
an additional set of tests. If it passes {any} of these further
tests, it is called a QSS. The oval labeled ``$H \approx 0$" 
represents a set of $7$ hierarchical conditions, while the oval
just below it represents a set of $3$ hierarchical conditions.  
We refer to any source selected by our algorithm,
whether it is an SSS or a QSS, as a ``very soft" X-ray source (VSS).
See \rd\ \& Kong 2003b for details.
}
\end{inlinefigure}
\vspace{-0.1cm}

\subsection{Previous Studies} 

We have used the algorithm to
conduct 
studies of QSSs and SSSs in a small number of external
galaxies. 
When we have had
enough counts for spectral fits,
the best-fit models for QSSs have tended to be
 significantly harder
than models expected for SSSs.

For example, in M31 we find $15$ SSSs and $18$ QSSs,
with $\sim 10\%$ of all XRSs falling into one of these $2$ categories
(\rd\ et al.\, 2003a).
Four of the VSSs are bright 
enough to permit spectral fits using
photons from a single image. One of these is a QSS and is clearly
harder than the other $3$, which are SSSs (Figure 2 of
\rd\ et al. 2003a). Each of the $3$ SSSs can be fit by a blackbody model;
the temperatures are 25 eV, 56 eV, and 122 eV.
The QSS is the only one of the four with a
significant number of photons evident out to $2$ keV; it 
 is fit by a combination of  a blackbody model
($k\, T = 89$ eV) and a power law model, 
($\alpha= 3.3$).   The central $8' \times 8'$ of M31 contains the largest 
fraction
of VSSs ($> 10\%$ of all XRSs), 
most of them are SSSs ($10$ SSSs out of $16$ VSSs). In contrast, the 
majority of M31 VSSs in the
outer disk are QSSs. 

In a study of very soft sources in 
M101, M83, M51, and NGC 4697, 
we found $77$ SSSs and $72$ QSSs (Di\,Stefano \& Kong 2003c).
Spectral fits could be carried out for $8$ individual SSSs; $7$
of these were well-fit by blackbody models with 
$50$ eV $\leq k\, T \leq  101$ eV, with a median value of $71$ eV.
Spectral fits could be carried out for $10$ individual QSSs; $7$
of these were well-fit by blackbody models with
$56$ eV $\leq k\, T \leq  222$ eV, all but one with $k\, T > 120$ eV,
and with a median value of $142$ eV.

In M104 (NGC 4594), the Sombrero galaxy, we found among $122$ XRSs in 
total, 
$22$ VSSs, only $5$ of which are SSSs (\rd\ et al. 2003b).
The dominance of QSSs relative to SSSs
in M104, which
is viewed edge-on, could be due to two effects.
First, absorption in the disk is so severe 
that we are primarily sensitive to VSSs located more than $100$ pc
above the disk, more  likely to be part of an older stellar
population. Second, absorption is more likely to 
obscure or alter the observed spectra of SSSs. 
One of M104's QSSs, X-35, appears to be associated with a globular
cluster (GC). No source with both comparable luminosity and spectral
properties has been discovered in the GC system of
either the Galaxy or M31. 
In NGC 4472, we found 211 X-ray sources; 5 are SSSs and 22 are QSSs.
We also found $6$ GCs associated with VSSs, $2$ SSSs and $4$ QSSs
(\rd\ et al. 2003c).

\subsection{This paper}

The goal of this paper is to
study the properties of QSSs in as much detail as possible, 
especially to elucidate spectral differences between QSSs and
both SSSs and canonical XRSs. We begin in \S 2 with a study of
the broadband spectral properties QSSs, SSSs, and canonical XRSs in six
galaxies whose VSS populations we have already 
identified through previous archival studies.  
The galaxies are M101 and M83, two spirals
viewed almost face-on; M51, an interacting galaxy; M104,
a spiral seen almost edge on, and two elipticals, NGC 4472, and NGC 4697.
These galaxies were chosen for the work described in \S 2
because we have the most detailed information about
their broadband spectra. In addition, we used a deeper \chandra\ 
observation of M51 (OBSID1622) in this paper.
As part of our earlier studies, we have determined the count rates in each 
of $8$ energy bins:  
$0.1-0.3$ keV,
$0.3-0.5$ keV, $0.5-0.7$ keV, $0.7-0.9$ keV, $0.9-1.1$ keV,
$1.1-1.5$ keV, $1.5-2$ keV, and $2-7$ keV. 

Because we aim to extend the study of QSSs to a larger
number of galaxies, spanning a greater variety of galaxy environments,
in \S 3 we discuss the analysis of archival data from
a total of $19$ galaxies, including M31 and the $6$ galaxies
discussed in \S 2. 
In addition to NGC 4472 and NGC 4697, we now consider $7$ 
other elliptical
galaxies, and an S0 galaxy.  We add to the $4$ spirals 
of \S 2, $5$ additional spirals
observed at a variety of orientations, and
located at a broader range of distances. 
To each galaxy, even those studied before,
we apply one single procedure to detect XRSs and
to select VSSs.  
We identify $19$ QSSs that
can be subjected
to spectral fits. Since some QSSs may be absorbed SSSs, this
last step is important to test whether
a significant subset are
genuinely harder. 

We consider a variety of physical models for QSSs in \S 4. We turn in
\S 5 to discussions of the size of QSS populations in galaxies.  In
this regard it is especially important to consider our own Galaxy.
Since our algorithm would identify few (if any) QSSs in the Milky Way,
we focus on the question of how to identify QSSs in the Milky Way.  We
present our conclusions, and consider the prospects for future
research in \S 6.

\section{Broadband Spectra}  

Here we sketch the procedures for source detection and the selection of VSSs
 detailed in the papers on these specific
galaxies:
M101, M83, M51, and NGC 4697 (\rd\ \& Kong 2003);
M104 (\rd\ et al. 2003b); and NGC 4472 (Friedman et al. 2003;
\rd\ et al. 2003c). 
Source detection was accomplished using WAVEDETECT, with the energy
range $0.1-7$ keV; each source was visually inspected and any spurious
sources were eliminated. We analyzed only data taken
with the S3 CCD. (The exception was NGC 4472, for which we analyzed the
data from all of the CCDs.)
{\it Chandra}
deep field studies were used to estimate the expected numbers of background
sources, $\sim 10$ per S3 field, given the typical exposure times; 
this is much smaller than
the numbers of sources detected.
We eliminated all XRSs associated with known foreground stars, which
can have spectra and timing properties similar to VSSs. In addition,
we tested for the expected foreground and background VSSs by
running our detection algorithm on data
from several extragalactic fields analyzed by the {\it ChAMP} team.
We found that, given typical exposure times, 
the S3 fields would typically contain $2-4$
VSSs.\footnote{In fields for which we have optical data, we find that
these are often foreground stars.}
 Because this number is small compared to the numbers of VSSs
actually detected, it is likely that the majority of VSSs
are actually associated with the galaxies themselves. Further
evidence
that most XRSs (including VSSs) are
associated with the galaxies comes from the fact that the majority
are clearly associated with galaxy features, such as
spiral arms, or exhibit density variations that appear related
to the distance from the galaxy centers. 

Table 1 summarizes the results on the relative sizes of the SSS, QSS,
and canonical XRS populations in each of M101, M83, M51, M104, NGC 4697, and
NGC 4472. (We did not include M31 because only a small portion of
the galaxy has been observed with S3, and the results differ from field
to field, as described in \S 1.1.)
The total number of sources detected is $762,$ of which $211$ 
($28\%$) are VSSs. Of the VSSs, %52\% ($122$ in total) are QSSs;
$89$ sources were identified as SSSs.    

Direct comparisons of the numbers of each type of
source from galaxy to galaxy can be misleading, since the
detection limits are different, and conditions within the galaxies,
such as the gas distribution, strongly influence SSS detection and
have not been studied.   
Nevertheless, two features do stand out.  First, each galaxy
has a significant population of VSSs, including QSSs.
While SSSs were expected, there is not a large pool of local 
sources (in the Galaxy
or Magellanic Clouds) known to have
 the spectral properties of QSSs and
luminosities in the range observed for these extragalactic sources,
typically above $10^{37}-10^{38}$ erg s$^{-1}$.   
Second, the ratio of the number of QSSs to SSSs appears to be larger
for the two ellipticals and for M104.
Since both ellipticals are along
directions
with low-$N_H,$ and because absorption by gas along the line of
sight within these galaxies is less likely, it seems probable that the QSSs
they contain are  not simply absorbed SSSs, but sources with intrinsically
harder spectra. Furthermore, because the stellar populations of the 
ellipticals are likely to be old, Table 1 may indicate that QSSs
are more common, compared with SSSs, among older populations. This would
be consistent with  the relative dominance of QSSs over SSSs
in (1) M104, and in (2) regions of M31 near the outer portions of the disk,
where little or no recent star formation has taken place.

Figure 2 shows the percentage of the photons from
all SSSs (bottom panel), QSSs (middle panel), and canonical XRSs
(top panel) 
with energies in each of the $8$ energy bins: 
$0.1-0.3$ keV,
$0.3-0.5$ keV, $0.5-0.7$ keV, $0.7-0.9$ keV, $0.9-1.1$ keV,
$1.1-1.5$ keV, $1.5-2$ keV, and $2-7$ keV. 
This figure confirms the selection procedure
is working. SSSs, for example,   
have little or no emission above $1.5$ keV.  
QSSs are significantly harder than any of the SSSs
detected in the Galaxy and Magellanic Clouds, in that a significant fraction
($\sim 30\%$) of the radiation they emit is in the form of photons with
energies above $1.1$ keV. 
These sources will not be chosen by procedures designed to
select SSSs. Yet, they  are  
significantly softer than canonical XRSs, since photons
with energies above $2$ keV are either absent from their spectra
or constitute only a small fraction ($< 5\%$) of the total emission.  
Thus, the sources we have dubbed QSSs, based on the properties
of their broadband spectra, seem to span the spectrum between
SSSs and canonical XRSs.
An interesting feature is that only 
roughly $30\%$ of the photons from so-called
canonical XRSs 
have energies above $2$ keV  (see also \rd\ 2003).

\begin{table*}
\caption{Summary of sample galaxies}
\footnotesize
\begin{tabular}{lcccccccccccc}

\hline
\hline
Name& Type& Distance & $D_{25}$ & $M_B$ & Inclination & $N_H$ &
{\it
Chandra} & Total & VSS & SSS & QSS & Canonical \\
    &     &  (Mpc)   &  (arcmin)&       &             &
($10^{20}$cm$^{-2}$) & exposure (ks)& &
 & &&\\
\hline
M101 & Sc & 5.4, 6.7$^1$ & 23.8 & -20.45 & $0^{\circ}$ & 1.2 &
94.4& 118 & 53 &32 & 21 & 65\\
M83 &  Sc & 4.7, 4.57$^2$ & 11.5 & -20.31 & $24^{\circ}$ & 3.8 &
49.5& 128 & 54 & 28 & 26 & 74\\
M51  & Sc\,$^b$ & 7.7, 8.4$^3$ & 13.6 & -20.75 &
$64^{\circ}$ & 1.6 &
28.6& 92 & 36 & 15 & 21 & 56\\
M104 & Sa & 8.9$^4$ & 8.4 & -22.98 & $79^{\circ}$ & 3.7 & 18.5 &
122 & 22 & 5 & 17 & 100\\
NGC 4697 & E & 11.7$^6$, 15.9$^5$, 23.3 & 7.1 & -21.67 &
$44^{\circ}$\,$^c$ & 2.1 & 39.3 & 91 & 19 & 4 &15 & 72\\
NGC 4472 & E & 16.8 & 11.4 & -21.82 & $35^{\circ}$\,$^c$ & 1.7 & 34.4 &
211 & 27 & 5 & 22 & 184\\

\hline

\end{tabular}
\par
\medskip
\begin{minipage}{0.95\linewidth}
\footnotesize

Notes---All data are from Nearby Galaxies Catalogue (Tully
1988)
unless specified.\\

$^a$ Ratio of number of SSSs to total number of X-ray sources.\\
$^b$ Seyfert 2 galaxy.\\
$^c$ For elliptical galaxies
the inclination is given by $3^{\circ}$ +
acos\,$\left(\sqrt{((d/D)^2 - 0.2^2)/(1 - 0.2^2)}\,\right)$, where
d/D
is the axial ratio of minor to major diameter (Tully
1988).
This inclination angle is generally unrelated to the value of
$N_H$.\\

Referrences---1: Freedman et al. 2001; 2: Karachentsev et al. 2002;
3: Feldmeier et al. 1997; 4: Ford et al. 1996; 5: Tonry et al.
2001; 6: Faber et al. 1989
\end{minipage}
\par

\end{table*}

\vspace{.2 true in} 
\begin{inlinefigure}
\psfig{file=f2.ps,height=3in,angle=-90.0}
\caption{
Broadband summed spectra of
SSSs, QSSs, and canonical XRSs in M101, M83, M51, M104, NGC 4697, and NGC 
4472.
% 77 SSSs (bottom panel), 110 QSSs
%(middle panel), and 241 
%canonical X-ray sources (top panel). 
The percentage of all the photons found in each of $8$ energy bins
(see text)
is plotted versus the energy.
%Top panel: all sources; bottom panel: sources with $< 50$ counts, but $> 25$ counts. 
%SSSs: dash-dot curve; QSSs: solid dark curve; canonical XRSs: dashed curve.
}
\end{inlinefigure}
\vspace{-0.1cm}

Figure 3 is a color-color diagram which shows the relative positions
of QSSs (filled green circles), SSSs (open red circles), and
canonical XRSs (blue triangles). (See Prestwich et al. 2003, for
this choice of axes.) All sources shown on this plot provided
more than $25$ counts, so that the hardness ratios are
fairly well defined.
Three features of this diagram
are worthy of note.
First, while there is some overlap along the boundaries, QSSs, SSSs,
and canonical XRSs each occupy distinct regions of the plane.
(Note that only the SSSs
are chosen by conditions that rely primarily on hardness ratios.
In other cases, the relative significance of the detection
in different bins is, e.g., also used.)

Second, examining the position of the QSSs,
relative to PIMMS (AO-6) predictions for a range of models (black curves),
we see that many QSSs are located in regions that are not associated with
$k\, T < 100$ eV models. While the QSSs which hug the left-most
line could be highly absorbed SSSs, QSSs lying to the right of
this line are more likely to have spectra described by higher
temperature blackbodies ($100$ eV $ < k\, T < 350$ eV),
or by a combination of such a model and a power law, with $\alpha$
softer than roughly $3.5$.

\vspace{-0.1 cm}
\begin{inlinefigure}
\epsfig{file=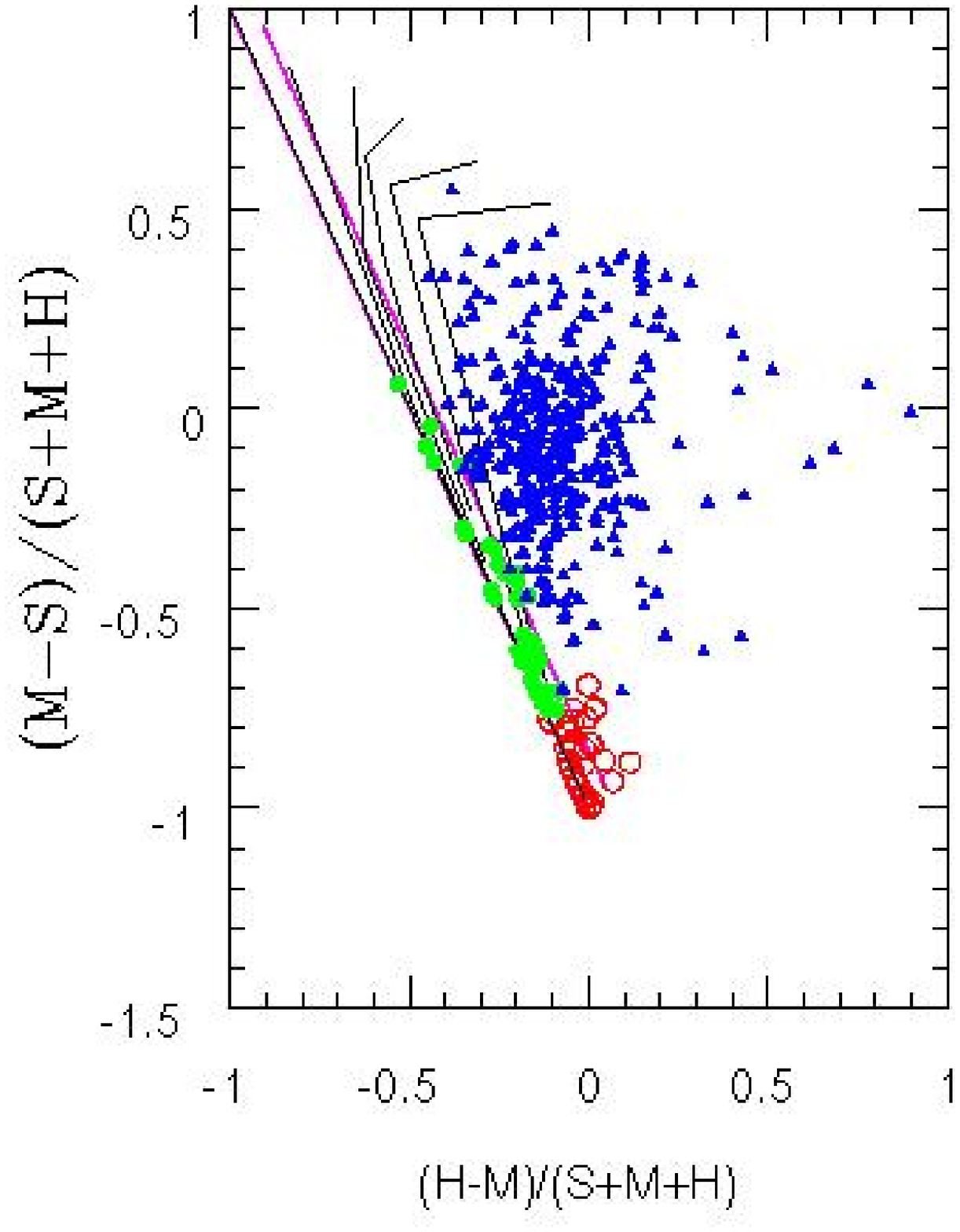,height=4.5in}
\caption{
All XRSs providing more than $25$ counts, located in one of the
$6$ galaxies listed in Table 1.  
SSSs: red open circles; QSSs: closed green circles; canonical XRSs:
blue triangles.
The two magenta lines correspond to theoretical curves that can be
used to identify SSSs and QSSs (see text). Each black curve
represents a physical model, with $N_H$ increasing
from the bottom ($N_H \sim 10^{20}$ cm$^{-1}$) 
toward the top ($N_H \sim 10^{22}$ cm$^{-1}$).
The left-most black curve is a pure blackbody (BB) model with $k\, T = 100$ eV.
 Moving rightward, successive curves are characterized by the following
sequence of models: BB, $k\, T = 175$ eV; BB, $k\, T = 250$ eV; BB, 
$k\, T = 300$ eV; power-law; $\alpha=3.3$; a composite model which
includes a power-law, $\alpha=3.3$ component, and a smaller
($20\%$ of the power-law flux) BB, with $k\, T = 300$ eV.  
While the QSSs that hug the left-most black line
could be absorbed SSSs, it is clear that many QSSs could be
associated with spectra that are intrinsically harder than
standard SSS spectra. 
}
\end{inlinefigure}
\vspace{-0.1cm}

Finally, the diagram illustrates that, at least for sources
providing enough counts that their position in a color-color
diagram is meaningful, it is possible to use hardness ratio conditions
to reproduce the overall effects (but not the full classification)
of our algortihm. All SSSs, e.g., have $(M-S)/(S+M+H) < -0.7.$  
This is not unique to SSSs, however, as some QSSs satisfy the same
condition. All QSSs lie between the $2$ magenta lines.
The left-most magenta line is defined by $H=0$, and can be
satisfied by an absorbed SSS. The second magenta line, parallel 
to the first, is defined by $H=(S+M)/20$
Thus, the combination of $3$ lines: the two magenta lines
described above, and the horizontal line at $(M-S)/(S+M+H)=-0.7$,
can be used to identify high-count SSSs and QSSs. For sources
providing fewer counts, however, the hardness ratios are not
well defined.
The algorithm therefore has an advantage,
in that it can select very soft sources
 in a way that is
 not subjective,   
 even if they provide
small numbers of counts.

%\subsection{Expectations}

%cycle 6; acis-s
%350 eV, 1e-12, 1e21
%.2-1.1 6.511E-02 
%1.1-2  9.761E-02
%2-7    1.478E-02 

%300 eV, 1e-12, 1e21
%.2-1.1 6.058E-02  
%1.1-2  9.034E-02 
%2-7    8.663E-03  

%alpha=3, 1e-12, 1e21
%.2-1.1 3.977E-02   
%1.1-2  1.992E-02  
%2-7    5.040E-03   

\section{Spectra of QSSs from $19$ Galaxies}

\subsection{Source Detection and Selection}

Data from 19 nearby ($D < 25$ Mpc) galaxies observed by \chandra\ with
the ACIS-S detector (see Table 2) were extracted from the \chandra\ Public Archive,
and further processed to remove times of background flares, using
standard CIAO background-filtering procedures
(http://cxc.harvard.edu/ciao/threads/filter\_ltcrv/). Data from the S1
chip (S2 if S1 was not selected) were used to generate background
lightcurves. Only data from the S3 chip were used for source analysis.

Sources were detected by running the CIAO celldetect tool on data in
two energy bands, $0.1 - 1.1$ keV, and $0.1 - 7$ keV, using a 
signal-to-noise
threshold of 2.5. Spurious sources, typically detected near detector
boundaries, were rejected manually. The two source lists were merged,
selecting the higher signal-to-noise detection %(NEED TO CHECK THIS) 
in
cases of multiple detections. Circular source and annular background
region files were generated for each source, with the source
region corresponding to a 90\% encircled energy fraction, determined
from the source off-axis angle.
% and the PSF size at an energy of
%0.6(2.5) keV for sources detected in the 0.1-1.1(0.1-7) keV band.
Data from near-by overlapping source regions were excluded from all
source and background regions. Net counts were extracted in three
energy bands, S: 0.1-1.1 keV, M: 1.1-2.0 keV, and H: 2.0-7.0 keV, using 
SAORD funtools utilities (http://hea-www.harvard.edu/RD/funtools/).

We applied our algorithm to the set of all detected sources to identify
SSSs and QSSs. To assure uniformity, our treatment of data from all $19$
galaxies was  
identical, using no information from previous studies.
The results are shown in Table 1. 

Because prior studies had been carried out for several
of these galaxies, we were able to compare our results
with those of previous studies, and find generally good agreement.

\subsection{Spectral Fits}

From the pool of all QSSs, we chose those   
providing more than $50$ counts, visually inspecting each
to
ensure that it is not in a region with significant
diffuse emission. This precaution lessens the possibility that
a soft source actually represents a dense unresolved region of diffuse
emission.
(See, e.g., Immler et al.\ 2003, and Figure 6 in Soria \& Wu 2002.)
Channels with 
energies less than 0.3 keV and greater than 7.0 keV were excluded.
The response files were corrected for the continuous degradation of
the ACIS detector quantum efficiency. The spectrum for each
source was fitted within XSPEC Version 11.2.0 with a simple
absorbed blackbody, multi-color disk blackbody, and
power-law models.  
For sources with 
more than 200 counts, we implemented fits using binned spectra (at least 
15 counts per spectral bin),
and employed  
the $\chi^2$ statistics to estimate goodness of fit. 
For fainter sources (50 $<$ counts $<$
200), we used unbinned spectra and applied the CASH 
(Cash 1979) statistics, employing Monte-Carlo
simulations to estimate the significance level. 
%The CASH statistics 
%is associated with a 
%maximum likelihood method designed to estimate the best-fit parameters 
%for unbinned data. It  is particularly useful for  sources that yield 
%few photons. 
%
%We justified the best fit of bright sources by 
%computing the reduced 
%$\chi^2$, while for faint sources, we performed Monte-Carlo 
%simulations to estimate the significance level of the fits. 

Table 2 lists the results.
For $7$ QSSs we fit binned spectra.
Sources 2, 7, and 8 
were well-fit by blackbody models, with
$k\, T$ ranging from $120$ eV to $220$ eV and  
X-ray luminosities from $1.4 \times 10^{37}$ erg s$^{-1}$ 
to $8.9 \times 10^{38}$ erg s$^{-1}$.  
Source 19 has an acceptable fit when a multicolor disk model is used,
but no acceptable blackbody model fits; $k\ T$ is $350$ eV, and $L_x$ is 
$1.7 \times 10^{39}$ erg s$^{-1}$. Although there is no single well-accepted
boundary defining ultraluminous X-ray sources (ULXs), the value 
$10^{39}$ erg s$^{-1}$ is often used. By this definition 
source 19 is a ULX. 
Source 6 was fit neither by a blackbody nor multicolor disk model; instead
a power law ($\alpha = 3.8$; $L_x = 2.6 \times 10^{38}$ erg s$^{-1}$) 
was acceptable, but this model required additional
Gaussian line emission at $1.27$ keV; a similar result was
found by Soria \& Wu 2003.
Sources 1, in M31 (see also Di Stefano et al. 2003a) and 18,  
in NGC 1399, were fit by power-law models, with
($\alpha = 3.6, L_x = 3 \times 10^{37}$ erg s$^{-1}$;
$\alpha =  2.9, L_x = 2.3 \times 10^{39}$ erg s$^{-1}$, respectively). 
%It is not clear that this last source 
%should have been picked out by an algorithm to 
%identify very soft X-ray sources. 
The ranges of temperatures and power-law indices are similar
for the sources fit using unbinned spectra.

\begin{table*}
\begin{center}
TABLE~2

{\sc Spectral fits to the bright QSSs}
\footnotesize
\begin{tabular}{llccccccccc}

\hline \hline
&&&    & \multicolumn{3}{c}{Blackbody} & \multicolumn{3}{c}{Power-law} &\\
&Galaxy (OBSID)$^a$& Distance &Source & $N_H$ & $kT$/$kT_{in}$ &
$\chi^2_{\nu}$/dof,
MC$^b$ &$N_H$ &  $\alpha$ & $\chi^2_{\nu}$/dof, MC$^b$ & $L_X$ (0.3--7
keV)\\
 &      & (Mpc) &(CXOU)& ($10^{21}$ cm$^{-2}$) & (keV) & &($10^{21}$
cm$^{-2}$) & & &($10^{38}$ erg s$^{-1}$)\\
\hline
1&M31 (1575)& 0.78 $^1$&J004244.3+411607 & --- &--- & ---
&$1.6^{+0.5}_{-0.3}$ &
$3.6^{+0.3}_{-0.3}$ & 1.0/45 & 0.3\\
2&M33 (786)& 0.84 $^2$ &J013341.9+303849 & $3.3^{+2.3}_{-1.4}$ &
$0.12^{+0.03}_{-0.03}$
&1.7/13 &---&---& --- & 0.14\\
3&M83 (793)& 4.7 &J133702.4-295126 & $2.4^{+2.5}_{-1.3}$ &
$0.15^{+0.03}_{-0.02}$ & 0.97&
$6.5^{+3.5}_{-2.4}$& $6.7^{+2.3}_{-1.6}$&0.71 & 0.5, 38\\
4&& & J133706.0-295514 & $1.5^{+1.7}_{-1.5}$ & $0.16^{+0.05}_{-0.03}$
&0.97
& $4.2^{+2.7}_{-2.3}$ & $5.4^{+3.5}_{-1.6}$ & 0.78& 0.26, 4.6\\
5&& & J133717.2-295153 & $1.1^{+1.1}_{-1.1}$ & $0.23^{+0.05}_{-0.04}$
&0.97
& $5.7^{+2.7}_{-2.1}$& $5.5^{+1.6}_{-1.3}$& 0.92 &   0.26, 11.6\\
6&& &J133649.0-295258 & --- & ---& --- &$2.0^{+1.4}_{-1.1}$&
$3.8^{+0.8}_{-0.7}$$^c$ & 0.9/12 & 2.6\\
7&M51 (354)& 7.7& J132939.9+471236 & 0.2 (fixed) & $0.22^{+0.03}_{-0.02}$
&1.16/13 &---& ---& --- & 3.3\\
8&M104 (1586)& 8.9 &J124001.1-113723 & $0.09^{+0.46}_{-0.09}$ &
$0.18^{+0.02}_{-0.02}$& 0.71/15 & ---& ---& --- &8.9\\
9&NGC4697 (784)& 15.9 $^3$ &J124834.4-055014 & 1 (fixed) &
$0.30^{+0.05}_{-0.04}$ & 1.0 & $1.0^{+1.1}_{-0.9}$ & $2.5^{+0.8}_{-0.6}$
& 0.7 & 2.5, 3.8\\
10&NGC4649 (785)& 16.8 &J124341.1+113248 & 1 (fixed)&
$0.22^{+0.04}_{-0.03}$ & 0.99 &$2.4^{+2.0}_{-1.6}$ &  $3.3^{+1.2}_{-1.0}$
& 0.46 & 2.1, 6.8\\
11&&&J124341.3+113320 & 1 (fixed) &
$0.26^{+0.04}_{-0.04}$ & 0.92& $0.32^{+0.28}_{-0.21}$ &
$3.9^{+1.6}_{-1.2}$  & 0.58& 2.6, 14.9\\
12&NGC3379 (1587)& 9.5 $^2$ &J104747.9+123445 & 1 (fixed)&
$0.25^{+0.04}_{-0.04}$ & 0.98 & $2.2^{+2.4}_{-1.7}$ & $3.6^{+1.4}_{-1.1}$
& 0.47 & 1.1, 3.8\\
13&&&J104741.8+123745 & 1 (fixed) &
$0.19^{+0.01}_{-0.02}$ & 1.0 &$2.0^{+1.3}_{-0.9}$ & $3.9^{+0.9}_{-0.7}$ &
0.81& 3.8, 12.9\\
14&NGC3115 (2040)& 11 $^4$ &J100510.0-074530 & 1 (fixed) &
$0.25^{+0.03}_{-0.03}$
& 1.0 &
$0.7^{+1.1}_{-0.7}$ &$2.7^{+0.8}_{-0.5}$  & 0.36& 2.1, 3.0\\
15&NGC1399 (319)& 20.5 $^5$ &J033826.7-352704 & 1 (fixed) &
$0.29^{+0.06}_{-0.04}$ & 0.89 & $2.9^{+2.0}_{-0.5}$  &$3.5^{+1.1}_{-1.0}$
& 0.64& 2.5, 10.0\\
16&&&J033832.3-352645 & 1 (fixed) & $0.17^{+0.02}_{-0.02}$ &1.0&1 (fixed)
& $3.8^{+0.4}_{-0.5}$& 0.93& 3.8, 6.1\\
17&&&J033831.7-353058 & 1 (fixed) &
$0.18^{+0.01}_{-0.03}$ & 1.0 &$0.13^{+0.52}_{-0.13}$ & $2.6^{+0.6}_{-0.3}$
& 0.73 & 6.7, 5.5\\
18&&&J033827.6-352648 & ---& ---& ---&$0.9^{+0.9}_{-0.3}$ &
$2.9^{+0.5}_{-0.4}$ & 1.1/22 & 23.4\\
19&&&J033831.8-352603 & 0.15 (fixed) & $0.35^{+0.03}_{-0.04}$$^d$
& 1.0/26& ---&--- &---& 16.6\\
\hline
\end{tabular}
\end{center}
\footnotesize
NOTES --- All quoted uncertainties are 90\% confidence. Distance is quoted 
from Nearby Galaxies Catalogue (Tully 1988) unless specified.
Sources without reduced $\chi^2$ and dof values were fitted with
unbinned data using CASH statistics; ``goodness-of-fit'' 
determined by Monte-Carlo simulations (MC) is employed and parameters of 
both blackbody and power-law models are shown. The luminosities refer to 
blackbody 
(first) and power-law (second) model. \hfill \\
{\bf $^a$} OBSID is the observation identification number of \chandra\ 
observation. For the nine remaining galaxies that were used in this paper 
but do not appear in this table, the OBSIDs are as follows: M32 (2017), 
M101 (934), NGC4258 (1618), NGC4472 (321), NGC4552 (2072), M84 (803), 
NGC415 (348), NGC5845 (4009), and NGC1313 (2950).% \hfill \\
{\bf $^b$} For faint sources, we list the probability that the best fit model 
would produce a lower value of the CASH statistic than that calculated 
from the data, as determined via XSPEC Monte-Carlo simulations.  A best 
fit model should have a value of about 0.5.% \hfill \\ 
{\bf $^c$} A 1.27 
keV Gaussian emission line is required to obtain a good fit. %\hfill \\
{\bf $^d$} Multi-color disk blackbody model. \hfill \\
References --- $^1$ Stanek \& Garnavich 1998; $^2$ Freedman et al. 
2001; $^3$ Faber et al. 1989; $^4$ 
Durrell et al. 1996; $^5$ Merritt \& Ferrarese 2001
\end{table*}

\section{Physical Models}

As is the case for SSSs, the empirical definition of QSSs is
broad, allowing the class to possibly encompass a wide range
of physical phenomena. For SSSs, the key clue was provided by the
effective radius and the fact that hot WDs, whether the central stars
of planetary nebulae,
post-novae, or quasi-steady nuclear burners, were expected to
have temperatures as hot as those observed for SSSs. It is therefore
not surprising that more than half of the local systems appear to be
WD systems of known types.  
Nevertheless, the physical natures of the local candidates, 
such as CAL 83 and CAL 87, for the
less-well established NBWD models are
not yet definitely determined. 
Neutron star and BH models cannot be
ruled out. The spatial distributions of SSSs in
face-on spirals (see \rd\ \& Kong 2003a ,c) indicates that a
significant fraction may be young systems,
with ages less than a few times $10^8$ years; this is
not expected for most NBWD models.  
Indeed, 
two SSSs in M31 
appear to be associated with SNRs (Di\,Stefano et al. 2003a).   
Nevertheless, the large majority of SSSs for which we have
multiple observations are highly variable or even transient,
ruling out SNR  models. Some SSSs are ultraluminous, most
likely ruling out WD models and suggesting the possibility that they
are accreting intermediate-mass BHs (IMBHs). 
The effective radii are also compatible with  IMBH models (see below).

Just as the SSS moniker can likely corrspond to several different
physical models, so can the label ``QSS".   
As for SSSs, we use the effective radius to obtain a clue
as to the nature of the sources.
\begin{equation}
R_{eff}=695 km\, \Big({{L}\over{10^{38} {\rm erg s}^{-1}}}\Big)^{{1}\over{2}}
                  \Big({{200\, eV}\over{k\, T}}\Big)^2
\end{equation}
This is smaller than the radii of WD, but larger than the radii of 
NSs. It is roughly $23$ times larger than the Scharzschild radius
of a $10\, M_\odot$ BH, or roughly $8$ times larger
than the inner disk radius, if there is ongoing accretion extending
inward to the last stable orbit.
Judging simply by the effective radius, 
the most natural model would therefore be one
in which the accretor is a BH with mass larger than
expected for the remnants of present-day stars.
(In \S 4.2 we show that this conclusion is supported even 
for the standard disk models used for BH accretors.) 
Models other than those invoking an IMBH are 
nevertheless possible, and we therefore 
begin by considering them in \S 4.1, 
before focusing on IMBH models in \S 4.2.

\subsection{Comparisons with well-studied systems}

\noindent{\sl White Dwarfs: }  
For QSSs, $R_{eff}$ 
 is less than half the radius of the smallest WD 
(one with $M= 1.4\, M_\odot$). In fact, if we assume that
the surface of a $1.4\, M_\odot$ WD is radiating at the Eddington luminosity,
we find that the maximum possible temperature is $\sim 150$ eV.  
Calculations of emission from hot WDs which are either
experiencing or have recently experienced 
nuclear burning find, however, that the photosphere is located
well above the surface, and temperatures tend to be below $100$ eV
(Nomoto 1982; Iben 1982).
Thus, while it may be possible for QSS-behavior to be
exhibited by a WD, this would require that the emission emanate from
a limited portion of the WD surface, or else that there be
a significant amount of reprocessing of the soft radiation
emitted from the photosphere. Behavior like this has not been 
observed in those local SSSs either known or suspected to be hot WDs.
We note, however,
that some QSSs from which we have recieved only a small number of photons
are classified as QSSs rather than SSSs on the basis of
just a few photons with energies above $1.1$ keV. 
Deeper observations of such sources might reveal them to be SSSs,
or might find that the 
energy in photons above $1.1$ keV is small enough 
to be consistent with modest upscattering of 
photons from a hot WD, or with shocking due to winds, etc.
Only roughly $1/4$ of, e.g., the VSSs of Table 1 provide fewer than $20$ counts,
however, so such an explanation is unlikely to apply to the majority of QSSs. 
In summary, although they cannot be ruled out,
 WD models are not the preferred models for most QSSs.
   
\smallskip
 
\noindent{\it Neutron Stars: } 
$R_{eff}$ is larger than NS radii. Nevertheless, when SSSs
with even larger effective radii were discovered, it was 
noted that, on theoretical grounds, NSs can
have extended photospheres  
(Kylafis \& Xilouris 1994).
This model was not the subject of much additional attention
as an explanation of SSSs, possibly because there seemed to be no
obvious mechanism to preferentially select WD-like radii.
If, however, some QSSs are NSs, their discovery may signal
that a continuum of effective radii is realized in nature,
eliminating the fine-tuning problem.     
In fact, at least
one NS systems with soft SSS-like emission and beamed hard  
emission is known. The X-ray pulsar 
RX~J~0059.2-7138 (Hughes 1994)  
has a luminous soft unpulsed component 
(see also Kohno, Yokogawa, and Koyama 2000). 
It is unlikely that RX~J~0059.2-7138 is an anomaly.
In fact, Her X-1, a Galactic X-ray pulsar, has long been
known to have a luminous time-variable soft component
(Shulman et al.\. 1975; Fritz \etal\ 1976). 
If this is common, then a set of soft sources in external
galaxies may simply be X-ray pulsars for which we are out of the
beam of the harder emission. RX~J~0059.2-7138 would
likely be identified as a SSS; it is conceivable, however,
that other X-ray pulsars have luminous soft components with
spectra like QSSs instead of SSSs.
Another conjecture is that some QSSs
and SSSs represent temporal states of NS 
 in which
the hard emission is off, but the soft emission continues.    
Such states could occur fairly frequently among local
X-ray binaries, and, since the {\it RXTE's All-Sky Monitor (ASM)}   
cannot detect the soft emission ($< 2$ keV) that would be dominant,
we would only be aware of
them if they happened to occur during specific 
observations with detectors having good soft sensitivity.  
In summary, some QSSs could be NS binaries from which we have not
detected hard radiation because of either the angle from which 
we view the system or the temporal state of the system.
In either case, the ability to detect and study QSSs in other
galaxies would have implications for population estimates and
evolutionary calculations; in addition,  we should be able
to identify a sizable population of local analogues    
(see \S 5.2). 

\smallskip 

\noindent{\sl Stellar-Mass Black Holes:} 
Analogous arguments for temporal states 
can 
be applied to stellar mass BHs.
In fact, locally-studied stellar mass BHs do exhibit
``thermal dominant" (TD), or ``soft" states. 
These states are, however,
much harder than QSS spectra, with typical 
values of $k\, T$ for the TD states of stellar-mass BHs
larger than 600 eV. 
For a recent detailed review, see McClintock \& Remillard (2004). 
During intervals when the inner
disk temperature appears to be lower, i.e., the inner disk radius
is larger, the radiation is not dominated by thermal
emission from a disk, but
rather by a 
a harder power-law component (see, e.g., Sobczak et al. 2000).
Even though stellar-mass BH spectra are harder than 
QSSs, they appear to be the the closest analogues among physical states that
are actually known to exist in our Galaxy. 
It would be interesting if some QSSs
are BHs, as there is no reliable way, except possibly through
characteristics of time variability associated with, e.g., X-ray novae 
(see, e.g., Williams et al. 2004 and 
references therein), to identify BHs in other galaxies. 
It is important to note, however, that the square of the
 temperature of the inner
disk is inveresly proportional to the mass of the central BH.
Larger mass BHs should exhibit softer spectra. If, therefore,  
there is not a sharp truncation in the BH mass function,
TD states with softer spectra are expected, and may provide
an explantion for QSSs.  

\smallskip

\noindent{\sl SNRs:\, }
The LMC contains several 
SNRs (N63A, DEM 71, 
N132D, and N49) with luminosities between $3\times 10^{36}$ erg s$^{-1}$ 
and $3 \times 10^{37}$ erg s$^{-1}$ have $0.6$ keV $< k\, T_{eff} < 0.8$ keV
(Hughes, Hayashi, \& Koyama 1998).
Thus, although they are cool compared with other bright SNRs, 
they are hotter than QSSs. Furthermore, on
theoretical grounds, is expected that the SNR should dim as is it cools. 
Note however, that we do find apparent counterexamples in the X-ray data 
from M31.  
Near the center of 
M31, there are three X-ray SNRs (identified by both X-rays and radio) that 
have temperatures of $\sim 200-250$ eV (Kong et al. 2002; Kong et 
al. 2003). One of these $3$ might be bright enough to have
been observed in more distant galaxies. 
Variability studies can identify QSSs that are {\it not} SNRs. 
In M31 (\rd\ et al.\, 2003), e.g., we find that both QSSs and SSSs 
tend to be variable over month-to-year time scales. We have also found 
that roughly half of the bright QSSs in M51 are variable.

As summarized above, each standard model, 
WD, NS, stellar-mass BH accretors, or SNRs, can potentially produce emission
with the range of temperatures and luminosities exhibited  by the
QSSs we have observed in other galaxies. No one model, however, 
appears to be 
an especially natural candidate to explain the majority of QSSs.

\subsection{Intermediate-Mass Black Holes}

It is possible that the relatively soft spectra of QSSs,
as compared with stellar-mass BHs, indicates that QSSs
are associated with BHs of somewhat higher mass. Indeed,    
if we assume that $R_{eff}$ should correspond to the radius
of the last stable orbit around a Schwarzschild BH, at $3$ times the
Schwarzschild radius, then 
\begin{equation} 
M= 77.2\, M_\odot \Big({{L}\over{10^{38} erg s^{-1}}}\Big)^{{1}\over{2}}
                  \Big({{200\, eV}\over{k\, T}}\Big)^2  
\end{equation} 
For the blackbody fit parameters in Table 1, the BH masses   
computed from this formula 
range from approximately $30\, M_\odot$ up to nearly $300\, M_\odot$,
with $6$ ($3$) masses greater than $100\, M_\odot$ ($200\, M_\odot$),
and a median value of $87\, M_\odot$. For a thin but optically thick
disk, the appropriate formula is identical to Eq.\, 2, but with a
factor included on the right-hand side to take into account the effects
of spin, spectral hardening, and orientation.
For Schwarzscild BHs, the factor
is typically $\sim 1.7-2,$ leading to larger mass estimates, 
as high as $\sim 600\, M_\odot$.
For spinning black holes, the spectral hardening parameter 
tends to be smaller, but there is an overall factor of $3$,
producing generally higher mass estimates.
Furthermore, the true luminosities may be significantly
 larger than the estimated X-ray
luminosities.  
In summary, QSS spectra are consistent with the IMBH hypothesis.

In fact QSS spectra can be viewed as extensions of the TD states
of stellar-mass BHs to softer disks and, therefore to possibly 
larger masses.  Furthermore, 
some of the QSSs may represent extensions
of ULX spectra to lower luminosities. Evidence for cool disks comes from
fits to several ULX spectra (see., e.g., Miller et al.\, 2003, Kong \&
\rd\ 2003;, Zampieri et al.\, 2004). Most recently,
Wang et al.\, (2004) have used Monte Carlo simulations to conduct a
self-consistent study of the disk and a Comptonized power law
component and found an inner disk temperature $< 250$ eV for $6$
of seven ULXs. Even the most luminous QSSs considered in \S 2 
provided far fewer photons than the ULXs studied by Wang et al. (2004).
Thus, since the power-law component is 
typically more than an order of magnitude dimmer than the 
thermal component (see also Mc Clintock and Remillard 2003),
it is conceivable that the sources we find to be predominantly soft do
have a harder component which provides too few photons 
to significantly influence the fit. 
 
Because $L/L_{edd} \sim 0.13\, [M/(100\, M_\odot)] [k\, T/(200\, eV)]^4$,
the X-ray luminosities of these QSSs tend to be $\sim 0.01$ of the Eddington
value, which can be compatible with optically thick thermal emission. 
We note further, that the X-ray flux
would be only a small portion of the energy output by the BH.
For example, only the region just around the inner edge of the disk
is hot enough to emit X-rays. 
Until recently, it was thought that soft states in BHs were predominantly
also ``high," luminous states. Observations of several systems have,
however, suggested that such states occur over a range of luminosities,
including relatively low (but not quiescent) states.  
(See Homan et al.\, 2001; Wijnands \& Miller 2002; 
McClintock \& Remillard 2004.)

If any of the ULXs now
considered as possible IMBHs in other galaxies (e.g. Miller \& Colbert 
2003),
actually are IMBHs, QSSs fit in as a natural
part of the class.   
First, with luminosities above $10^{39}$ erg s$^{-1}$,
 some QSSs may themselves be viewed as ULXs. 
Second, the QSSs that are not ULXs
may represent the lower $\dot M$ systems that should 
constitute the {\it bulk} of the IMBH population.
Indeed, if any of the ULXs currently under study
do harbor IMBHs, it is clear that there must be
a less luminous population of IMBHs, since, in the context
of most binary models, the least luminous IMBHs are expected to be the
most numerous representatives of the class.   
It has been convincingly demonstrated that the total energy output in some
ULXs is indeed large, consistent with the high
estimated X-ray luminosities (Pakull \& Mirioni 2002). Spectral studies 
of a small but growing number of systems suggest signs of cool disks 
expected for accreting IMBHs (e.g., Miller et al. 2003; Kong 
2003; Cropper et al. 2004; Miller et al. 2004). If 
some ULXs are IMBHs,
then it is certain that at least some
galaxies are home to large numbers of less luminous
IMBHs. The dominant spectral component for those IMBHs accreting at rates
 not too much below the
Eddington luminosity is expected to be soft.
It therefore seems possible that the spectral selection procedure
we have employed can identify, from among the many XRSs in a
galaxy, those most likely to be IMBHs. 
The accretion rates ($\dot M = L_x/(0.1\, c^2),$
for the systems in Table 1 range from
roughly $2 \times 10^{-9} M_\odot$ yr$^{-1}$ to
$4 \times 10^{-7} M_\odot$ yr$^{-1}$, consistent with binary models
(\rd\ 2004).

\section{Galactic Populations of QSSs}  

\subsection{QSSs in other galaxies} 

With the limited information we have so far, it is not
possible to compute the size of typical galactic populations
of QSSs. Because we do not yet know
what they are, we cannot be guided by theoretical
models. Furthermore, the data collected so far provide only limited information.
The detection limits vary significantly among the galaxies we
have studied, ranging from
$\sim 5 \times 10^{35}$ erg s$^{-1}$ for long ($\sim 50$ ksec)
observations of M31 to just under $10^{38}$ erg s$^{-1}$
for $20-50$ ksec observations of galaxies at $10-15$ Mpc.
(To compute these limits, we assume
that 14-20 counts are required to both detect a source and classify it
as QSS, and that $N_H \sim 10^{21}$ cm$^{-2}$; we use PIMMS AO6 release
to estimate count rates.)
We find sources down to the detection limits
of the galaxies we have studied in detail: 
M31, M83, M51, M104, NGC 4697, and NGC 4472.
Because of its proximity,
M31 may provide the best guide. 
In {\it Chandra} observations which are effectively sensitive to QSSs in 
only
roughly $1/5$ of the area of M31, e.g., we have identified $18$ QSSs.
We can therefore argue for at least several dozen QSSs in M31.

\subsection{QSSs in the Galaxy} 

If other galaxies harbor QSSs,
so must our own. In fact it is almost certain that some
detected XRSs would be identified as QSSs were they located
in other galaxies, such as M101. There are, however,  
few, if any obvious local candidates for the class.  
It is not surprising that we have apparently not yet identified
a pool of luminous sources with the spectral properties of QSSs in our Galaxy
or in the MCs. 
First, absorption in the Galactic plane 
modifies the spectra of soft sources significantly. 
Second, as described below, the surveys best poised to study Galactic sources
would have had some difficulties in studying QSSs.  
For example, a $10^{36}$ erg s$^{-1}$ source $5$ kpc from
us would yield a flux of $3.4 \times 10^{-10}$ erg s$^{-1}$ cm$^{-2}$.
If $k\,T = 200$ eV, the count rate for {\it ROSAT}'s PSPC would have been 
roughly $1$ count s$^{-1}$, and the source would have been easily detected
in the {\it All Sky Survey} ({\it RASS}). 
More than $90\%$ of the incident photons would
have energies above $1$ keV, however; with the PSPC's 
sensitivity cut off of $\sim 2.5$ keV,
there would have been no way to determine that there is little flux
at higher energies. 
Thus, while the difference between SSSs and canonical XRSs was
readily detected by {\it ROSAT}, the difference between QSSs and canonical XRSs
was not.

In recent years, the most extensive and detailed information collected
on local X-ray binaries has been collected by {\it RXTE}.
Its sensitivity, however, does not extend below $2$ keV. {\it RXTE}
 can therefore not detect 
SSSs, or even those QSSs which emit no radiation in the $H$ band.
It is very likely, however, that 
the high-energy tails of some QSSs have been detected by {\it RXTE}. 
The measured flux would typically be on the order of a percent
of the bolometric flux.  
This suggests that  
QSSs may be selected by comparing the {\it RASS} source list 
with lists from
other catalogs of Galactic X-ray binaries (e.g. Liu, van 
Paradijs, \& van den Heuvel 2001) 
and carrying out spectral fits for 
{\it RASS} sources that (1) exhibit high count rates, but (2) have
low flux above $2$ keV,  as measured from fits in other catalogs
(e.g. Liu, van
Paradijs, \& van den Heuvel 2001), 
and which (3) are not identified with foreground stars
or background galaxies. 

We note that IMBH models for some QSSs are attractive because, in them,
QSSs represent a natural extension of  known systems.  First, the class of QSSs
can be viewed as a possible soft extension of the TD states of well-studied
stellar-mass BHs. Second, QSSs, and possibly some SSSs, with luminosities
below $10^{39}$ erg s$^{-1}$ can be viewed as possible extensions
to lower luminosity of some ULX candidates for IMBHs. Indeed, since
systems accreting close to the Eddington limit are rare,
QSSs of the type studied in this paper may
form the dominant population of accreting IMBHs.  

Existing {\it Chandra} and {\it XMM} data should also contain   
QSSs in deep exposures taken toward the Galaxy's center (Muno et al.\, 2003),
and in data taken during pointed observations and  archived as part of 
{\it ChaMP} (see Green et al.\, 2003) or 
{\it ChaMPlane} cf. Grindlay et al.\, 2003).
A broadband selection procedure can be developed to select QSSs
from among the other XRSs in these data sets. In order to be effective,
it would require a larger number of 
energy bins above $1.1$ keV than we have used so far.
A combination of 
broadband selection and spectral fits should then be able to identify QSSs.
If the density in and near the Galactic plane is roughly $1$ per cubic kpc,
we might expect to find roughly a dozen within a few kpc of Earth.

Identifying these systems would be important, because we 
can collect from them enough X-ray photons for good spectral
fits. Perhaps even more important, we can  
hope 
to obtain optical IDs for some nearby systems, 
detemining the natures of the  donor stars, measuring orbital parameters,
studying disk properties at other wavebands, etc. 
In this way we may be able to determine 
the physical natures of some QSSs. Even one well-studied system
will be significant; but it will be particularly important to
develop a sample of QSSs, as it seems unlikely that
all of them can be described by a single physical model.

\section{Conclusions and Prospects}   

When we attempted to identify SSSs in other galaxies,
we found that each galaxy harbors, in addition to SSSs, a large population
of somewhat harder sources.
QSSs and SSSs are found in elliptical and spiral
galaxies, near galaxy centers,
in and near the arms of spiral 
galaxies, in galaxy halos (\rd\ \& Kong 2003d), 
and in globular clusters (\rd\ et al. 2003b; \rd\ et al. 2004).   
Whether there is a smooth continuum
between the softest and hardest sources, or whether there are
preferred ranges within the continuum remains to be seen.

QSSs are XRSs whose spectra fill the gap
between SSSs and luminous canonical X-ray sources.
The challenge raised by their discovery is to understand what
physical states they represent.  
With the same sort of broad definition used for SSSs,
the class of QSSs is likely to include a variety of
physical systems. Whatever the nature or natures of the sources, however,
it seems clear that   
QSS behavior is introducing us to 
new regions of the parameter space.  
If, e.g., some QSSs are hot WDs, then we must understand (1) why 
the emitting regions are small, or (2) how a large portion 
of the soft radiation is upscattered, or (3) if there is a 
warm plasma or other emitting region associated with the system.
In what way are WD QSSs different from the local WD systems?

If, on the other hand, QSSs are NSs, there are other sets of issues
to be explored. What determines the range of photospheric
radii? 
Are some of the soft sources X-ray pulsars, for which we are out of the 
path of the beamed emission?
Can physical principles be used to predict the 
relative numbers of NS systems emitting soft or hard radiation?

Finally, 
any QSSs that are accreting BHs raise different questions.
Can we use the discovery of QSSs to identify BHs in
distant galaxies? 
What fracion are IMBHs? 
Are the QSSs we have discovered in the GCs of M104 and NGC 4472
BHs with masses above that of typical stellar-mass BHs? If so,
do they represent the BHs predicted to exist at the center of
some star clusters?

In summary, 
the discovery of QSSs should inspire
theoretical work on possible hard states of hot WDs, and possible
 soft states of luminous 
SNRs, NS accretors, and BH
accretors across a range of masses.
Although we have argued that IMBH models are promising,
it is likely that several models are realized in nature.
The key point is that the generation of radiation in the QSS regime points to 
new sets of active states.
 
Longer observations of the galaxies in our study will
provide more detailed and reliable fits for the brightest sources and 
allow fits to be carried out for some of the dimmer QSSs. Additional
observations will also provide
checks for time variability, which can help distinguish between
X-ray binaries and SNRs and to determine the applicability
of stellar-mass BH models.
Searches for QSSs in our own Galaxy 
are nearly certain to yield individual systems close enough to be
subject to detailed study across wavebands, perhaps elucidating
the natures of the sources.

\begin{acknowledgements}
It is a pleasure to thank Jeffrey E. McClintock,
 Jon M. Miller, Andrea H. Prestwich,
  Frederick D. Seward, and Patrick Slane 
for conversations. 
This work was supported by NASA under an LTSA grant,
NAG5-10705.  
\end{acknowledgements}

\end{document}